\documentclass[twocolumn]{aastex6}

\newcommand{\kms}{km~s$^{-1}$}
\usepackage[hyphenbreaks]{breakurl}

\shortauthors{Redfield et al.}

\begin{document}


\title{Spectroscopic Evolution of Disintegrating Planetesimals: Minutes to Months Variability in the Circumstellar Gas Associated with WD\,1145+017}


\author{Seth Redfield\altaffilmark{1}}

\author{Jay Farihi\altaffilmark{2}}

\author{P. Wilson Cauley\altaffilmark{1}}

\author{Steven G. Parsons\altaffilmark{3}}

\author{Boris T. G\"{a}nsicke\altaffilmark{4}}


\and

\author{Girish Duvvuri\altaffilmark{1}}


\altaffiltext{1}{Astronomy Department and Van Vleck Observatory, Wesleyan University, Middletown, CT 06459, USA; sredfield@wesleyan.edu}
\altaffiltext{2}{Department of Physics and Astronomy, University College London, London WC1E 6BT, UK}
\altaffiltext{3}{Department of Physics and Astronomy, University of Sheffield, Sheffield, S3 7RH, UK}
\altaffiltext{4}{Department of Physics, University of Warwick, Coventry CV4 7AL, UK}

\begin{abstract}

With the recent discovery of transiting planetary material around WD\,1145+017, a critical target has been identified that links the evolution of planetary systems with debris disks and their accretion onto the star.  We present a series of observations, five epochs over a year, taken with Keck and the VLT, which for the first time show variability of circumstellar absorption in the gas disk surrounding WD\,1145+017 on timescales of minutes to months.  Circumstellar absorption is measured in more than 250 lines of 14 ions among ten different elements associated with planetary composition, e.g., O, Mg, Ca, Ti, Cr, Mn, Fe, Ni.  Broad circumstellar gas absorption with a velocity spread of 225 km~s$^{-1}$ is detected, but over the course of a year blue shifted absorption disappears while redshifted absorption systematically increases.  A correlation of equivalent width and oscillator strength indicates that the gas is not highly optically thick (median $\tau \approx 2$). We discuss simple models of an eccentric disk coupled with magnetospheric accretion to explain the basic observed characteristics of these high resolution and high signal-to-noise observations.  Variability is detected on timescales of minutes in the two most recent observations, showing a loss of redshifted absorption for tens of minutes, coincident with major transit events and consistent with gas hidden behind opaque transiting material.   This system currently presents a unique opportunity to learn how the gas causing the spectroscopic, circumstellar absorption is associated with the ongoing accretion evidenced by photospheric contamination, as well as the transiting planetary material detected in photometric observations.

\end{abstract}


\section{Introduction} \label{sec:intro}
A rapid improvement in our understanding of the formation and evolution of exoplanetary systems is taking place using several distinct regions
of the HR diagram.  At the earliest stages, the Atacama Large Millimeter/submillimeter Array (ALMA) and similar observations of protoplanetary disks are leading to insights in early chemistry and 
physical structure \citep{oberg15,vandermarel13}.  In mature, solar-type and low-mass stellar systems, {\em Kepler} transit observations and radial velocity studies are continuing to 
provide a bounty of exoplanetary systems in relatively close orbits, including numerous small and multi-planet systems \citep{vanderburg16}, but also true 
Jupiter analogs in colder orbits \citep{kipping14}.  State-of-the-art adaptive optics systems are providing giant planet and substellar companion detections 
in relatively wide orbits at intermediate-mass stars \citep{macintosh15,carson13}, and for which a few iconic systems appear to share some broad characteristics
of the solar system \citep{marois08,su13}.  These detections cover a broad range of phase space that constrain planet frequency, size and density as a
function of stellar mass and metallicity, but there still remain critical, empirical gaps in our knowledge.  For example, observational constraints on the interior structure of exoplanets are limited to model fits to the mean density \citep{madhu12b} and extrapolating from the observations of the outermost atmospheric layers \citep{redfield08brtr, millerricci10,madhu15}

Fortunately, nature provides a means to indirectly measure the bulk chemistry of planetary precursors or fragments of major planets, via polluted 
white dwarfs.  These ``retired" systems also provide insight into small planets and their assembly around intermediate-mass stars, which is not yet 
possible with any current or planned technique.  These advanced planetary systems are common.  Their frequency is at least 30\% \citep{koester14,zuckerman10},
providing a distinctive signature of exoplanetary systems \citep{rocchetto15,farihi13}.  While many exhibit
the observational hallmarks of gaseous and dusty debris from tidally-destroyed minor planets \citep{gaensicke06,farihi09}, due to the limited sensitivity of current infrared facilities to small planetesimal and disk masses \citep[see][]{wyatt14}, numerous systems only reveal their planetary nature via atmospheric metals \citep{xu12,bergfors14}.  
The atmospheric metal pollution itself enables the determination of the element to element abundances within the debris and disrupted parent body,
where the most detailed measurements have yielded chemistry that is remarkably similar to the bulk Earth \citep{gaensicke12,jura14}, along with diversities comparable to but beyond the meteoritic classes of the solar system \citep{xu13}, and with some outstanding objects that are rocky yet H$_2$O-rich \citep{farihi13b,
raddi15}.

Recent space and ground-based light curves of WD\,1145+017 provide striking confirmation of the canonical model with transits consistent with multiple, disintegrating planetesimal fragments.  In the first campaign of the {\it Kepler} K2 operations, WD\,1145+017 was identified and proposed by three groups to be observed (GO1071 PI: Redfield; GO1048 PI: Burleigh; GO1007 PI: Kilic).  The observations exhibit a complex transiting signature of multiple bodies orbiting on periods of 4.5--4.9 hours \citep{vanderburg15}.  This is the first white dwarf to be found to exhibit a transit signature.  Extensive ground-based campaigns have been performed \citep[e.g.,][]{gaensicke16, rappaport16}.  These photometric observations show a much higher incidence of transit activity than seen in the original K2 data.  A wide diversity of transit shapes are also detected, indicating substantial evolution of the system in only a matter of months.  

\citet{xu16} published high resolution spectra obtained by Keck.  These observations show strong photospheric lines indicative of recent accretion on to the star, as well as distinct absorption profiles caused by circumstellar gas.  The star exhibits a He-rich atmospheric spectrum punctuated by myriad, strong lines of heavy elements.  Given the rapid settling time for heavy elements ($<$ 10$^6$ yr, compared with $>$ 10$^8$ yr cooling age), the circumstellar material has been recently accreted onto the stellar surface.  
Low resolution ($\Delta\lambda \approx 1000$ \AA) spectroscopic observations show transit events akin to photometric observations, with no significant variation by wavelength \citep{alonso16}.  

The discovery of transiting events associated with WD\,1145+017 comes right on the heels of several white dwarf planetary systems exhibiting variability,
and implying dynamical evolution on human timescales.  The first variations of any kind were seen in the gaseous emission lines associated with the 
disk orbiting Ton\,345 \citep{gaensicke08}.  The gas component in the disk at SDSS\,J122859.93+104032.9 has persisted for at least 12 years and may 
vary periodically  \citep{manser16}, while similar emission from SDSS\,J161717.04+162022.4  has monotonically decreased, then disappeared over several years 
\citep{wilson14}.  Only within the past two years has variability also been witnessed via thermal emission in the infrared.  
A rapid drop in infrared emission has been seen in at least one system, SDSS\,J095904.69--020047.6, and it remains unclear how this might have 
occurred \citep{xu14}.  It has been suspected for some time that high-rate bursts of disk accretion take place onto polluted white dwarfs \citep{farihi12},
and the necessary, highly dynamical formation of the compact disks \citep{veras15} make such events likely, but as yet no variability associated with
changes in accretion rate or metal abundance have been confirmed.

This paper explores the variability and richness of high resolution spectra (five epochs over the course of a year) of WD\,1145+017, and emphasize the unique opportunity to measure the physical conditions (e.g., abundances, dynamics, etc...) in the circumstellar disk.  This so-far unique star provides an unprecedented window to observe planetesimal disintegration, catastrophic fragmentation, and disk evolution in real time.

\section{Observations and Data Reduction} \label{sec:obs}

\subsection{Keck / HIRES}
Observations of WD\,1145+017 were taken on 2015 November 14 using HIRES on Keck I \citep{vogt94} as part of program 15B/N116Hb (PI: Redfield), to monitor polluted white dwarfs.  The approximate resolving power of the observations is $R \equiv \lambda/\Delta\lambda \approx$ 34,000, or 8.8 \kms. The C5 decker was employed, which has a slit size of $7\farcs0 \times 1\farcs1$.  Two 20 minute exposures were taken at the end of the night at high airmass.  The signal-to-noise ($S/N$) per pixel of the extracted spectra is approximately 10 across much of the observed spectrum, which ranges from 3100--5890 \AA.

\begin{figure*}[!th]
\epsscale{.5}
\plotone{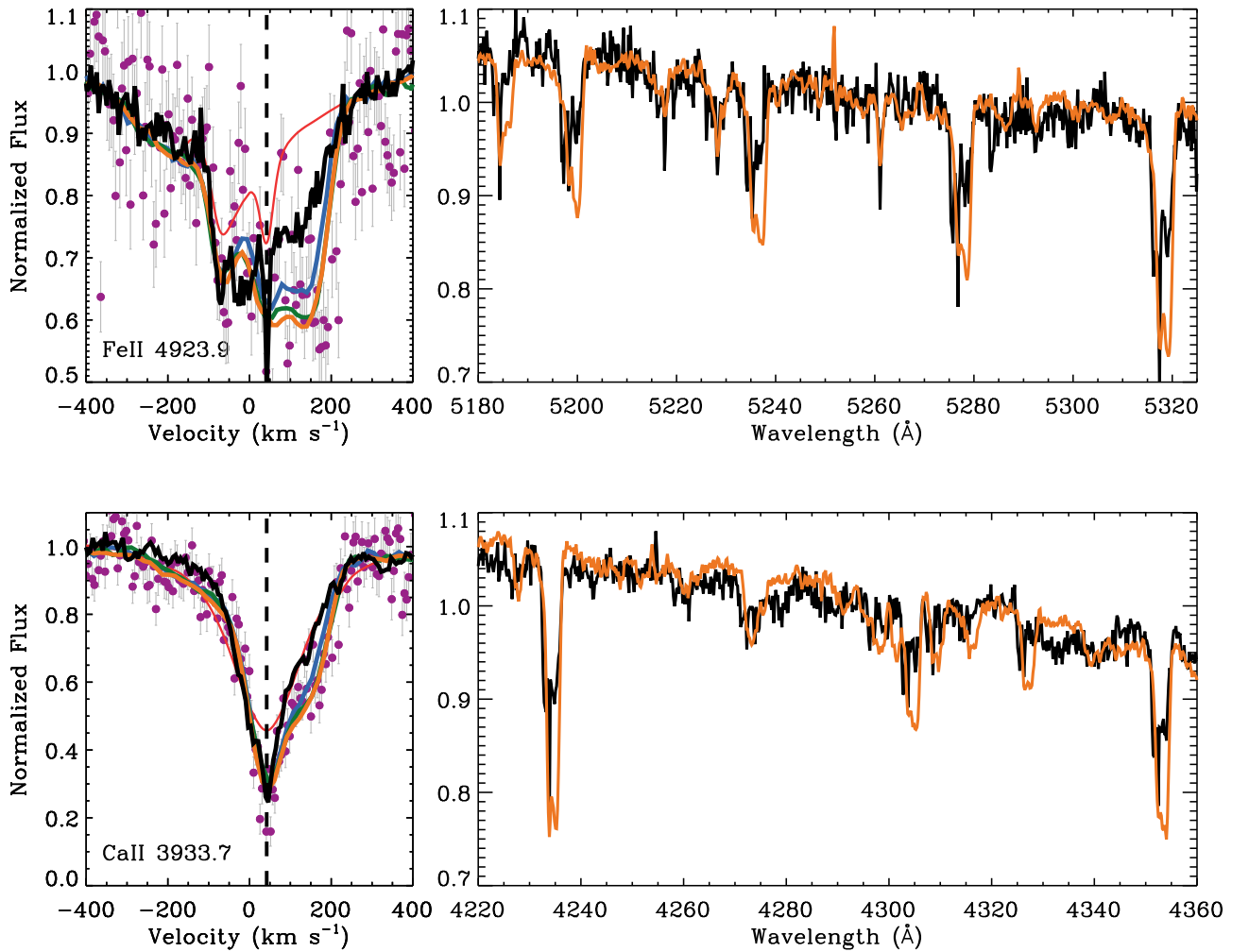}
\caption{({\it left}) Individual profiles comparing spectra obtained for five epochs (Keck/HIRES 2015 Apr [black] presented in \citet{xu16} and 2015 Nov [purple]  with 1$\sigma$ error bars, and VLT/X-shooter 2016 Feb [blue], 2016 Mar [green], and 2016 Apr [orange]).  A photospheric model is shown in red.  Clear circumstellar absorption is detected with 1) substantial variability on the red side of the feature, and 2) a diminishing of blue-shifted features observed in the earliest data. ({\it right}) Circumstellar features are ubiquitous in the spectrum.  A broader wavelength region is displayed with only the the two most distant observations in time, VLT/X-shooter Apr 2016 in orange and \citet{xu16} Keck/HIRES spectrum in black.  \label{fig:f1}}
\end{figure*}

The data were reduced using the HIRES Redux package written by Jason X. Prochaska\footnote{\url{http://www.ucolick.org/~xavier/HIRedux/}}. Standard reduction steps were taken, including bias subtraction, flat fielding, and the removal of cosmic rays and hot pixels. The spectra were extracted using a $6\farcs 6$ boxcar. All images were examined manually for order overlap in the blue chip. HIRES Redux also performs 2D wavelength solutions using ThAr lamp exposures taken at the beginning of each night. The residuals for the wavelength fits are $\approx$ 0.05 pixels, or $\approx$ 0.02 \AA, in all orders. When applied to an individual observation, all wavelength solutions are corrected for Earth's heliocentric velocity. 

We also make use of the Keck/HIRES spectra from program 15A/UCLA (PI: Jura) which were presented, and made available, by \citet{xu16}.  These are the earliest high resolution spectra available on this target, taken before the announcement of the K2 detection by \citet{vanderburg15}.

\subsection{VLT / X-shooter}

Observations of WD\,1145+017 were performed with X-shooter \citep{vernet11} on 2016 February 14, March 29, and April 8, under Director's Discretionary Time Program 296.C-5014 (PI: Farihi).  For the February run, the science target was placed on the slit and observed in nodding mode in all three instrument arms quasi-continuously for 5.1 hours while the star was above airmass 2.0.  In the UVB, VIS, and NIR arms respectively the exposure times were 300, 366, and 2 $\times$ 188 seconds.  Spectra were taken at four (ABBA) positions along the slit, separated by 5$^{\prime\prime}$, before re-setting the rotator to the parallactic angle.  In total there were 10 consecutive repetitions of this overall pattern, resulting in 40 separate exposures.  Acquisition and guiding were carried out in an attempt to minimize slit losses due to the current lack of the atmospheric dispersion corrector on X-shooter.  In order to increase $S/N$, the March and April runs were executed in stare mode for 2.9 and 5.0 hours, respectively.  The exposure times were 280, 314, and 240 seconds in UVB, VIS, and NIR arms respectively.  All observations were through slits of 0.8$^{\prime\prime}$, 0.9$^{\prime\prime}$, and 0.9$^{\prime\prime}$ for UVB, VIS, and NIR, respectively, resulting in nominal resolving powers of $R \approx$ 6200, 7450, and 5300.  The optical chips were read out binned by 2 pixels in the spectral direction and in low-gain and high-speed mode to minimize dead time.

The X-shooter data were reduced using the latest release of the X-shooter
Common Pipeline Library (CPL) recipes (version 2.6.8) within ESORex, the
ESO recipe Execution tool, version 3.12. The standard recipes were used
to extract and wavelength calibrate the data, the instrumental response
was removed using twilight observations of the standard star EG 274. We
found at optical wavelengths the additional flexibility of
optimally reducing each spectrum individually lead to better results
than the standard combination of ABBA nod sets. This is mainly due to
the non-negligible curve of the spectrum introduced by differential
atmospheric dispersion since the slit was not perfectly aligned at the
parallactic angle despite our attempts to minimize this effect. This
also led to relatively large slit losses at the shortest
wavelengths, i.e., $<$ 3200 \AA.

The optical spectral extractions performed in STARE mode were found to be clearly superior to those extracted in NOD mode, and thus only the former frames were used in the analysis.  The NIR arm was found to have insufficient signal for useful results; the 40 frames taken during the February nodded observations resulted in an overall $S/N$ $<$ 5 in the H band.

\section{Spectroscopic Features} \label{sec:features}

We detect photospheric and circumstellar lines in multiple transitions and ions of the major elements involved in terrestrial planet formation: primarily  Mg, Al, Si, Ca, Fe, and secondary trace elements such as Ti, Cr, Ni.  Figure~\ref{fig:f1} gives several examples of the circumstellar absorption profiles.  Two lines (\ion{Fe}{2} at 4923.9 \AA\ and \ion{Ca}{2} at 3933.6 \AA) are shown in detail on the left.  Larger segments of the spectrum are also shown comparing the first (2015 Apr; \citealt{xu16}) and last (2016 Apr) epochs, showing many examples of circumstellar absorption, mainly due to \ion{Fe}{2}.

\subsection{Photospheric \label{sec:photospheric}}

The photospheric metal abundances are measured by fitting white dwarf atmospheric models to the data with fixed $T_{\rm eff} = 15900$ K and $\log g = 8.0$, as estimated in \citet{vanderburg15}.  The model and spectral fitting procedure are detailed in \citet{koester10}.  Sharp photospheric lines are detected at a velocity of 42 km~s$^{-1}$, with 30 km~s$^{-1}$ associated with the gravitational redshift given the assumed bulk properties of this white dwarf ($M_\star \approx 0.6 M_\sun$ and $R_\star \approx 1.4 R_\oplus$), and therefore 12 km~s$^{-1}$ associated with the radial velocity of the star.  Note that modest changes in the mass or radius of the star, of order 10--20\%, result in significant changes in the gravitational redshift, and therefore also in the estimated radial velocity of the system.  The gravitational redshift could plausibly be anywhere between 25--45 \kms given reasonable values white dwarf mass (0.55--0.75 $M_\sun$).  However, this does not prevent accurate modeling of the spectra.  

The photospheric abundances in the new observations are consistent with the measurements by \citet{xu16} to within $<$ 0.2 dex for Mg, Al, Si, Ca, Fe, and the upper limit for C.  For Ti, Cr, and Ni there are many lines, but all seem to be affected by circumstellar absorption, so a definitive abundance measurement is difficult.  The only significant, though notable, discrepancy with the photospheric abundances presented in \citet{xu16} is that our O abundance is 0.2 dex lower.

Based on the published atmospheric metal abundances with modest updates
from the modeling presented here, the composition of the
debris accreted onto WD\,1145+017 can be inferred.  Using updated white dwarf
diffusion models\footnote{\url{http://www1.astrophysik.uni-kiel.de/~koester/astrophysics/astrophysics.html}} \citep{koester09},
both the early phase ($t < t_{\rm diff}$) and steady state ($t \gtrsim 5 t_{\rm diff}$)
abundances of the material can be calculated for O, Mg, Al, Si, Ca, Fe.

Interestingly, the material polluting the atmosphere of WD\,1145+017 is
broadly chondritic in O, Mg, Al and Si, but shows significant
enhancements in both Ca and Fe.  This is a curious mix of material that,
at least superficially, appears to be a mixture of crust-like and
core-like matter.  The abundance pattern also supports a modest excess
of oxygen in the accreted matter, with an excess of 45\% or 60\%, for the
steady state or early accretion phases, respectively.  The more
conservative numbers yield a water fraction by mass of 20\%.  If the 
trace hydrogen in the photosphere was delivered in bulk by a parent 
body with this water fraction, then the total (accreted) planetesimal 
mass would be $2.4 \times 10^{23}$ g or about one-quarter the mass of Ceres \citep{farihi13, raddi15}.  This value is in excellent agreement with the
total accreted mass as inferred from metals alone, which is $2.6 \times 10^{23}$ g based
on this work, and broadly agrees with previous estimates \citep{xu16}.

\subsection{Circumstellar}

Broad and distinctive circumstellar absorption is clearly present, which bracket and blend with the photospheric lines.  The circumstellar absorption is shifted by the systemic velocity of the system ($\approx$ 12 km~s$^{-1}$) but no significant gravitational redshift (relatively to the observer) is expected at its location.  We detect significant (i.e., $>$ 3$\sigma$ absorption measured in three or more epochs) circumstellar absorption in more than 250 lines from among ten elements and 14 ions (\ion{O}{1}, \ion{Na}{1}, \ion{Mg}{1}, \ion{Al}{1}, \ion{Ca}{1}, \ion{Ca}{2}, \ion{Ti}{1}, \ion{Ti}{2}, \ion{Cr}{2}, \ion{Mn}{2}, \ion{Fe}{1}, \ion{Fe}{2}, \ion{Ni}{1} and \ion{Ni}{2}).  Like optical spectra of the solar atmosphere, the ion that produces the most numerous features is \ion{Fe}{2}, with more than 100 transitions with detected circumstellar absorption.  The vast majority of detected transitions arise from low excitation energies ($<$ 4 eV).  Several ground state transitions are also detected and may be contaminated by absorption from the interstellar medium \citep{frisch11,johnson15}.  We see circumstellar absorption for \ion{O}{1} at 7775.4 \AA\, although no clear circumstellar absorption is detected for the hydrogen Balmer lines.  It is noteworthy that there is circumstellar absorption seen in \ion{O}{1}, but 
not in H$\alpha$.  While the models used here are not able to link these 
species to the total atomic abundance in gas, \ion{O}{1} is present at $> 6 \times 10^{-19}$ g~cm$^{-3}$, 
whereas for $n = 2$ \ion{H}{1} there is an upper limit of 10$^{-20}$ g~cm$^{-3}$. 

\begin{figure*}[!t]
\gridline{
\rotatefig{90}{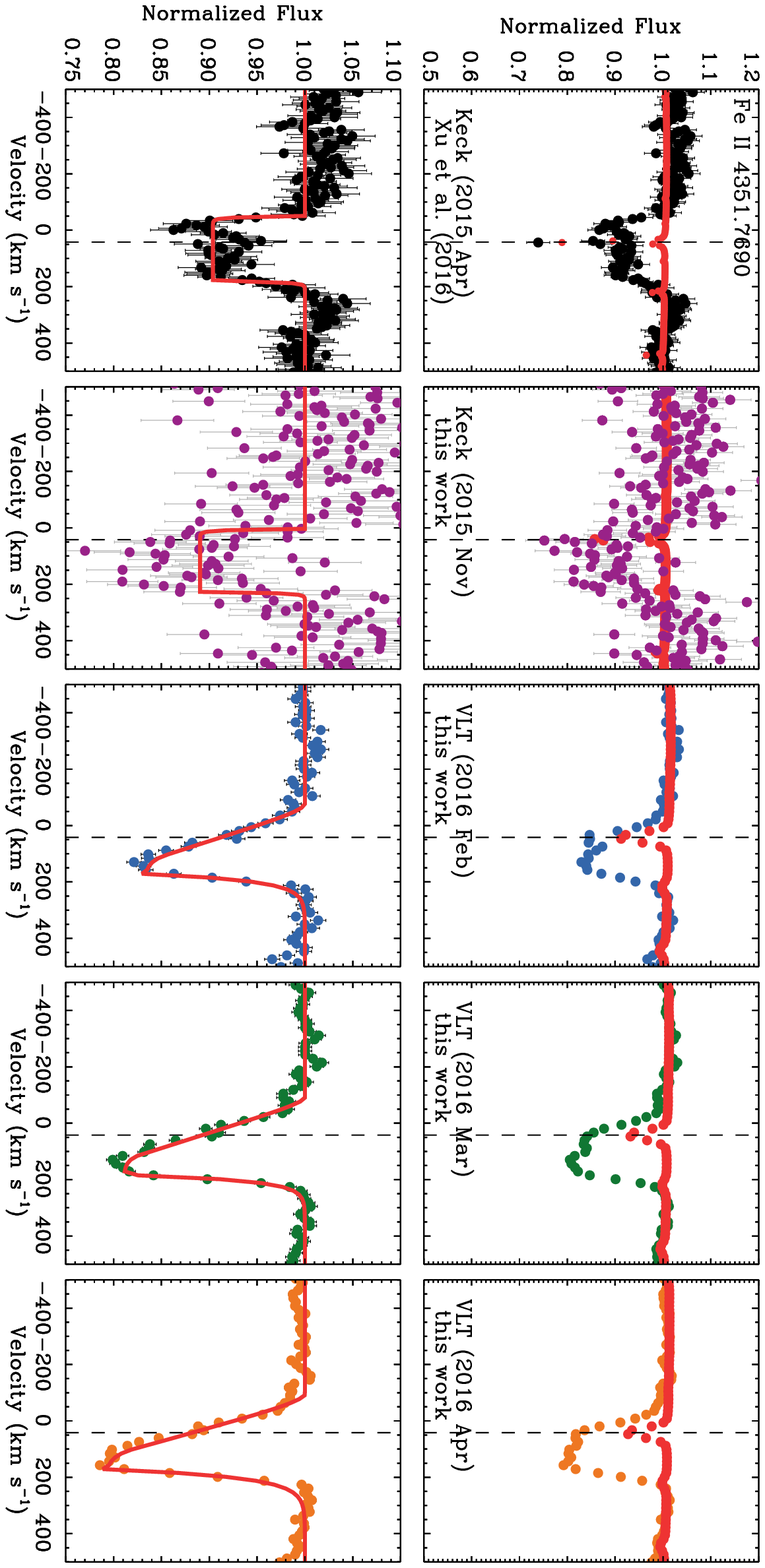}{0.38\textwidth}{}}
\gridline{\rotatefig{90}{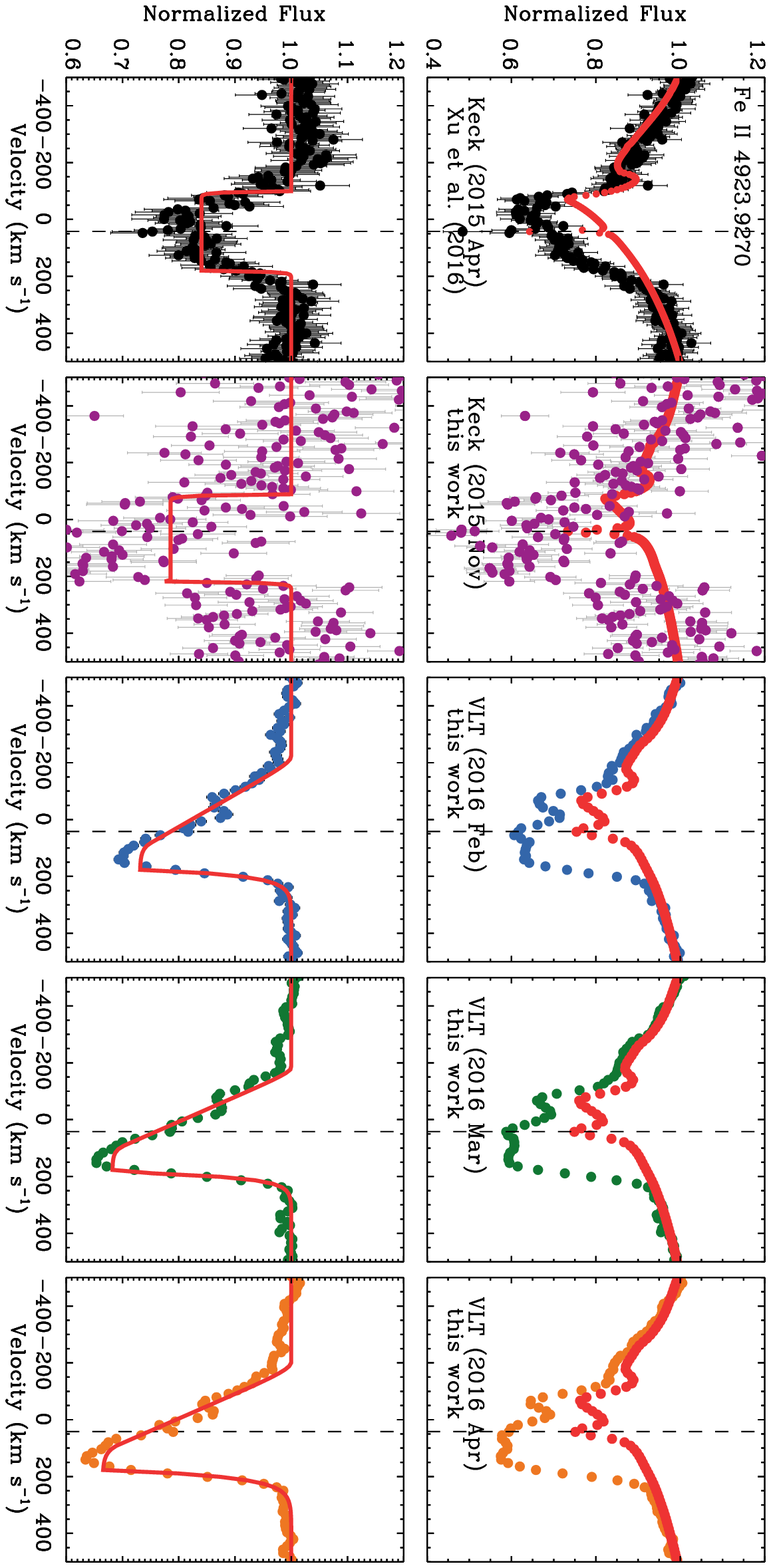}{0.38\textwidth}{}}
\gridline{\rotatefig{90}{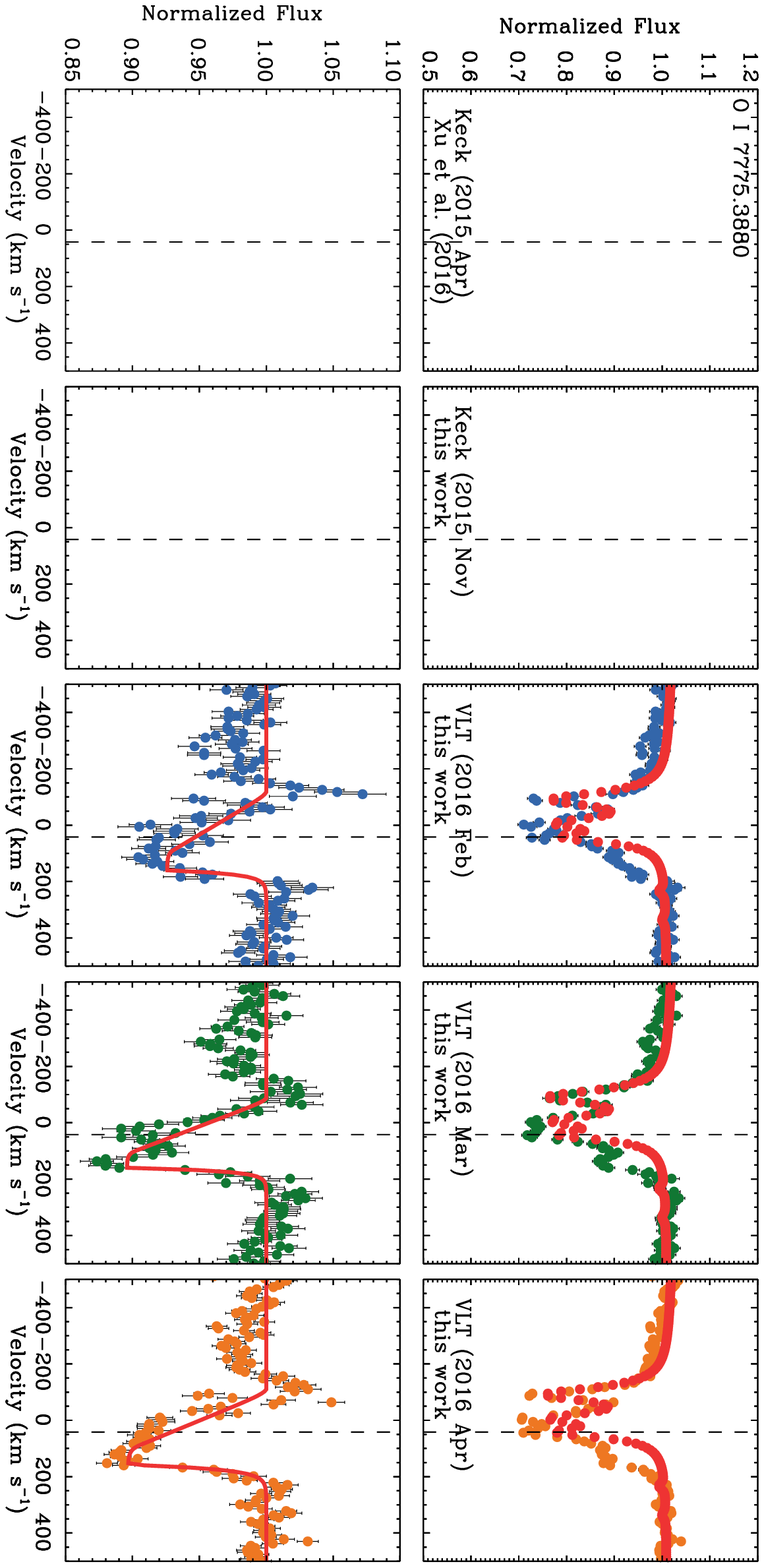}{0.38\textwidth}{}}
\caption{Examples of circumstellar profile fitting.  Three lines are shown: \ion{Fe}{2} 4351.7 \AA\ ({\it top}), \ion{Fe}{2} 4923.9 \AA\ ({\it middle}), and \ion{O}{1} 7775.4 \AA\ ({\it bottom}).  The top row for each shows the observed spectrum for each epoch with the stellar model overlaid in red.  The bottom row of each shows the stellar lines removed, leaving only the circumstellar absorption.  A simple trapezoidal model is shown in red (a simple box for the Keck epochs and a trapezoid with a tapered blue edge for the VLT epochs and convolved with the instrumental line spread function).  \label{fig:fits}}
\end{figure*}

\begin{figure*}[!t]
\includegraphics[angle=90,scale=.85]{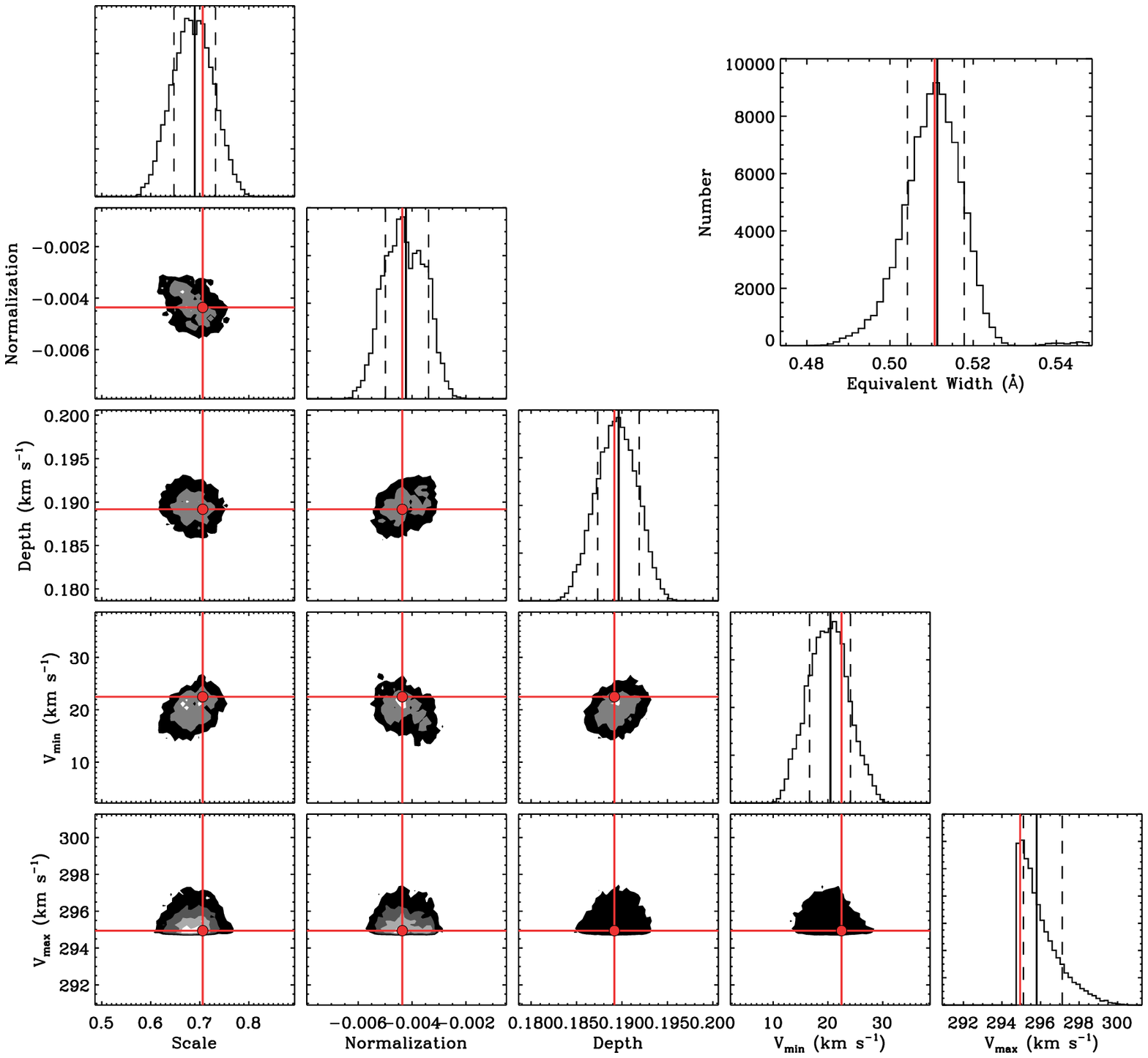}
\caption{Triangle plot showing the posterior probability distribution for the five parameters used in the circumstellar line fitting shown in Figure~\ref{fig:fits}.  This typical example is taken from the 2016 Mar epoch for the \ion{Fe}{2} 4351.7 \AA\ line.  The red lines indicate the values of the parameters at the best fit.  The solid black line indicates the median value and the dashed lines the 68\% confidence intervals.  The equivalent width posterior probability distribution is also shown in the upper right.  All distributions are well characterized . Note the distinctive asymmetry of the $V_{max}$ distribution, indicative of the extremely sharp velocity cutoff on the red side of the circumstellar absorption profile evidenced in the observed spectra.  \label{fig:mcmc}}
\end{figure*}

Table~\ref{tab:table} lists candidate identifications for all detected circumstellar features along with the basic details of the electronic transition.  We perform a Markov Chain Monte Carlo (MCMC) analysis on all detections.  After removal of the stellar lines, all circumstellar features were fit with a simple trapezoidal model (a simple box for the Keck epochs and a trapezoid with a tapered blue edge for the VLT epochs) in order to characterize the basic traits of the absorption.  Three parameters are used to create the trapezoidal model, a minimum velocity, a maximum velocity, and absorption depth.  Two additional parameters were fit to optimize the scaling of the stellar photospheric absorption removal and the normalization of the continuum.  The profiles are convolved with the appropriate instrumental line spread function.  A few examples of the fits to the circumstellar absorption are shown in Figure~\ref{fig:fits}.  A typical triangle plot for the MCMC analysis is given in Figure~\ref{fig:mcmc} for the 2016 Mar epoch of the \ion{Fe}{2} 4351.7 \AA\ feature.  The best fit values along with the median and 68\% confidence intervals are displayed.  The posterior probability distributions are well characterized.  Note the strong asymmetry in the maximum velocity parameter, signifying the sharp edge on the redward side of the observed circumstellar absorption profile.  The MCMC analysis provides a quantitative description of the absorption parameters and their errors.  The equivalent width is calculated from this fit and given in Table~\ref{tab:table}.  The full table has fits to all lines and gives the median values and 68\% confidence intervals for all parameters (e.g., minimum and maximum velocity and depth of absorption).  Due to the high density of circumstellar lines, blends are common, and complicate the fit values.  Blends are noted in the table, while isolated features give robust fits.

\section{Circumstellar Absorption Variability}
Observations over five epochs makes it clear that the circumstellar absorption is varying over month-long time scales.  

\subsection{Monthly Variability}

\begin{figure*}[t]
\epsscale{0.9}
\plotone{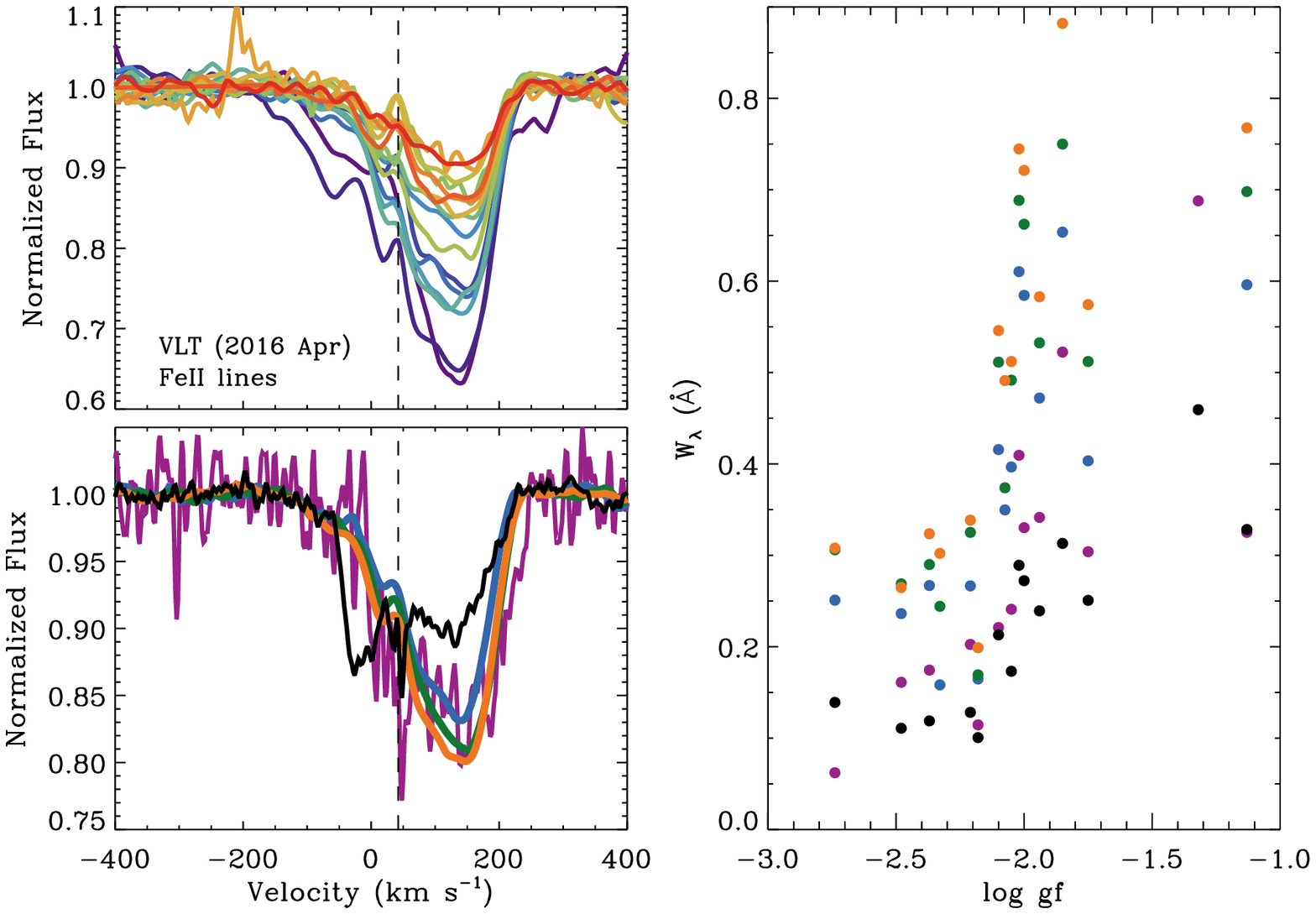}
\caption{({\it top left)} Comparison of 16 \ion{Fe}{2} circumstellar absorption features observed during the 2016 Apr epoch, where the color scale is proportional to the oscillator strength.  ({\it bottom left}) Mean profiles created by combining all the individual lines shown at top for all five epochs, where the color is the same as in Figure~\ref{fig:f1}.  As seen in the individual profiles in Figure~\ref{fig:f1}, significant changes are seen in the circumstellar absorption profiles.  ({\it right}) Equivalent width versus oscillator strength for \ion{Fe}{2}.  The colors correspond to each epoch, as in the plot on the bottom left.  The correlation indicates that the gas is not highly optically thick, although saturation is present at the largest oscillator strengths.  \label{fig:fe2comb}}
\end{figure*}

In order to evaluate the detailed line profile and variation between epochs, Figure~\ref{fig:fe2comb} displays a sequence of \ion{Fe}{2} lines that are strong, unblended, and relatively isolated form other features.  The sequence of transitions are shown overlaid in the top left for the 2016 Apr epoch.  For all epochs a mean profile is produced and overlaid in the bottom left panel.  The basic properties are 
\begin{enumerate}
\itemsep-0.1em
\item a broad absorption feature that is roughly 225 km~s$^{-1}$ in width
\item a range of depths up to 35\%
\item a constant and sharp maximum velocity of $\approx$ 210 km~s$^{-1}$ in the early epochs and $\approx$ 270 km~s$^{-1}$ in the late epochs, relative to the heliocentric rest frame
\item a minimum velocity of $\approx$ $-5$ km~s$^{-1}$ in the early epochs and $\approx$ 40 km~s$^{-1}$ in the late epochs, relative to the heliocentric rest frame.  
\end{enumerate}
The long term (i.e., monthly) evolution of the absorption tends toward deeper absorption on the red side of the circumstellar feature over time, while the maximum velocity increases slightly.  The strong blue-shifted circumstellar absorption (between --50 and 0 km~s$^{-1}$) observed in the early epochs has disappeared, and while the minimum velocity remains similar, the tapered appearance of the most recent data make this difficult to accurately measure.  This distinct change in the blue side of the circumstellar absorption line shape occurs between 2015 Nov and 2016 Feb.  In our simple trapezoidal fitting of the line profile, the two 2015 epochs are fit with a simple rectangular shape, whereas the 2016 epochs are fit with a trapezoid with a tapered slope on the blue side.


Aside from the changes in the spectral line shape, the total absorption (i.e., equivalent width) is systematically increasing with time.  The last three epochs increase 15--20\% from month to month.  As the transiting source material clearly continues to evolve, there is likely a continuous deposition of material that is feeding and growing the disk of gas that is responsible for this circumstellar absorption.  The absorption is not extremely optically thick, with a median optical depth $\tau \approx 2$.  As shown in Figure~\ref{fig:fe2comb}, there is a correlation in the equivalent width with oscillator strength.  

\subsection{Momentary Variability}

\begin{figure*}[!t]
\gridline{
\rotatefig{-90}{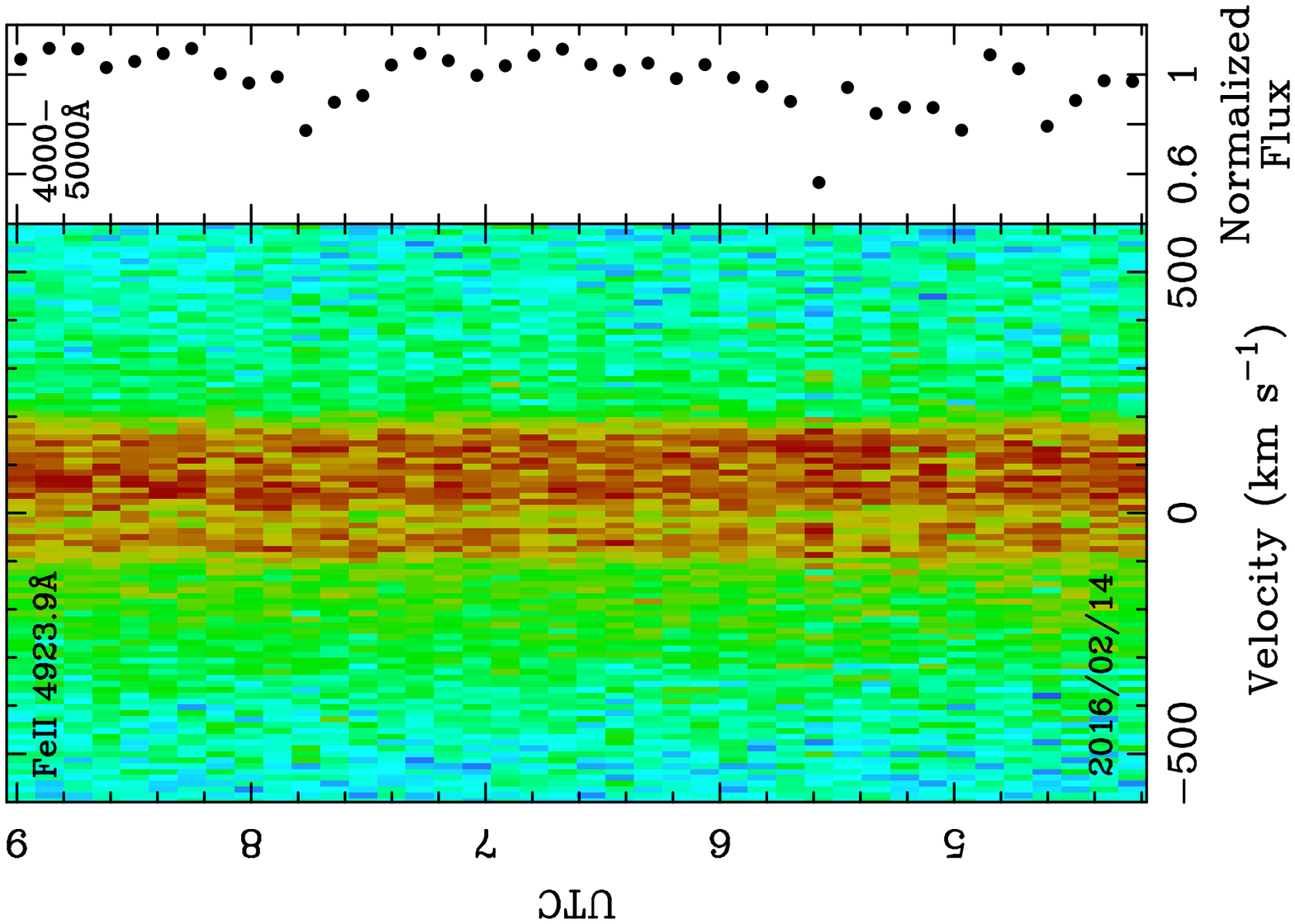}{0.47\textwidth}{}
\rotatefig{-90}{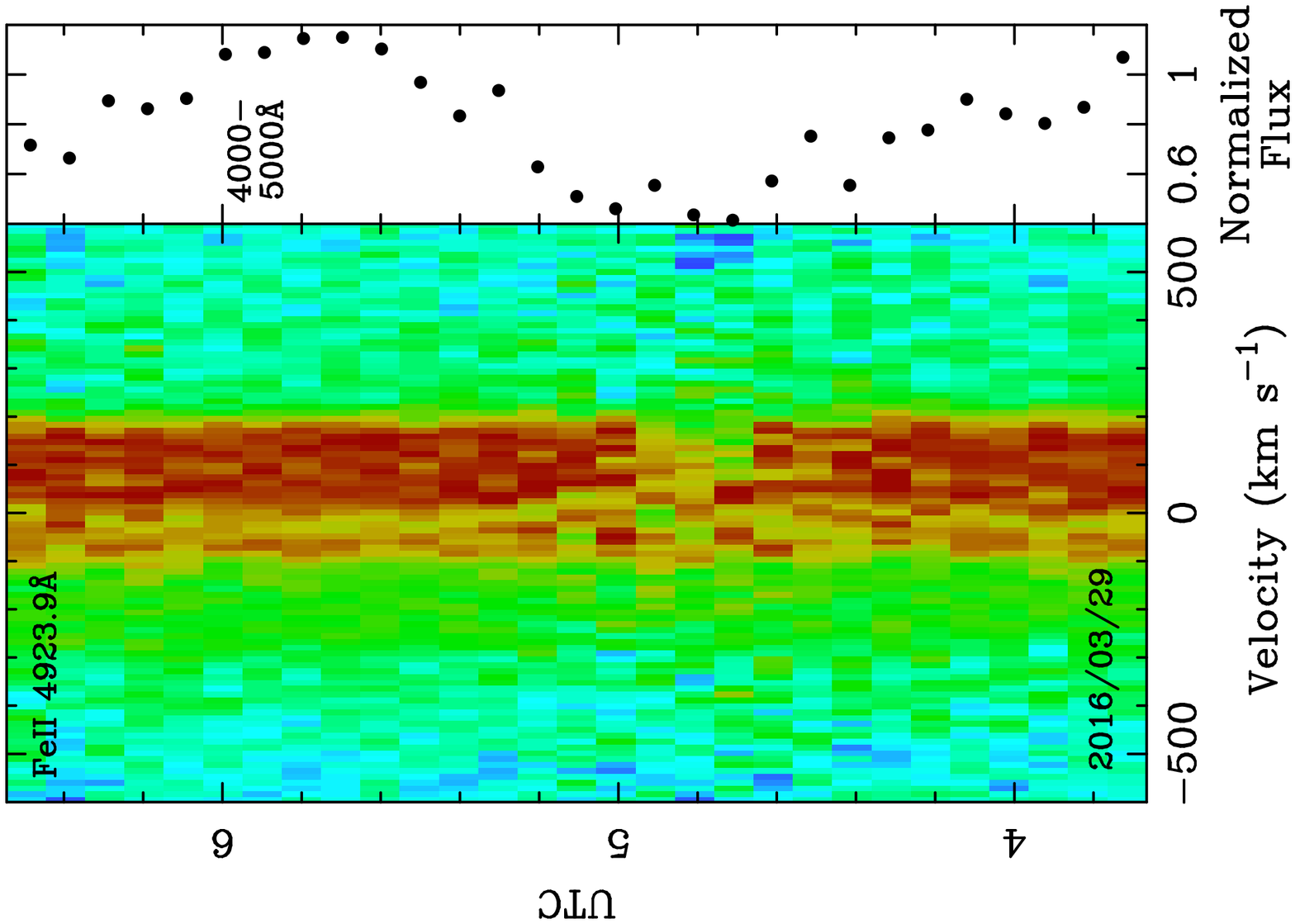}{0.47\textwidth}{}
\rotatefig{-90}{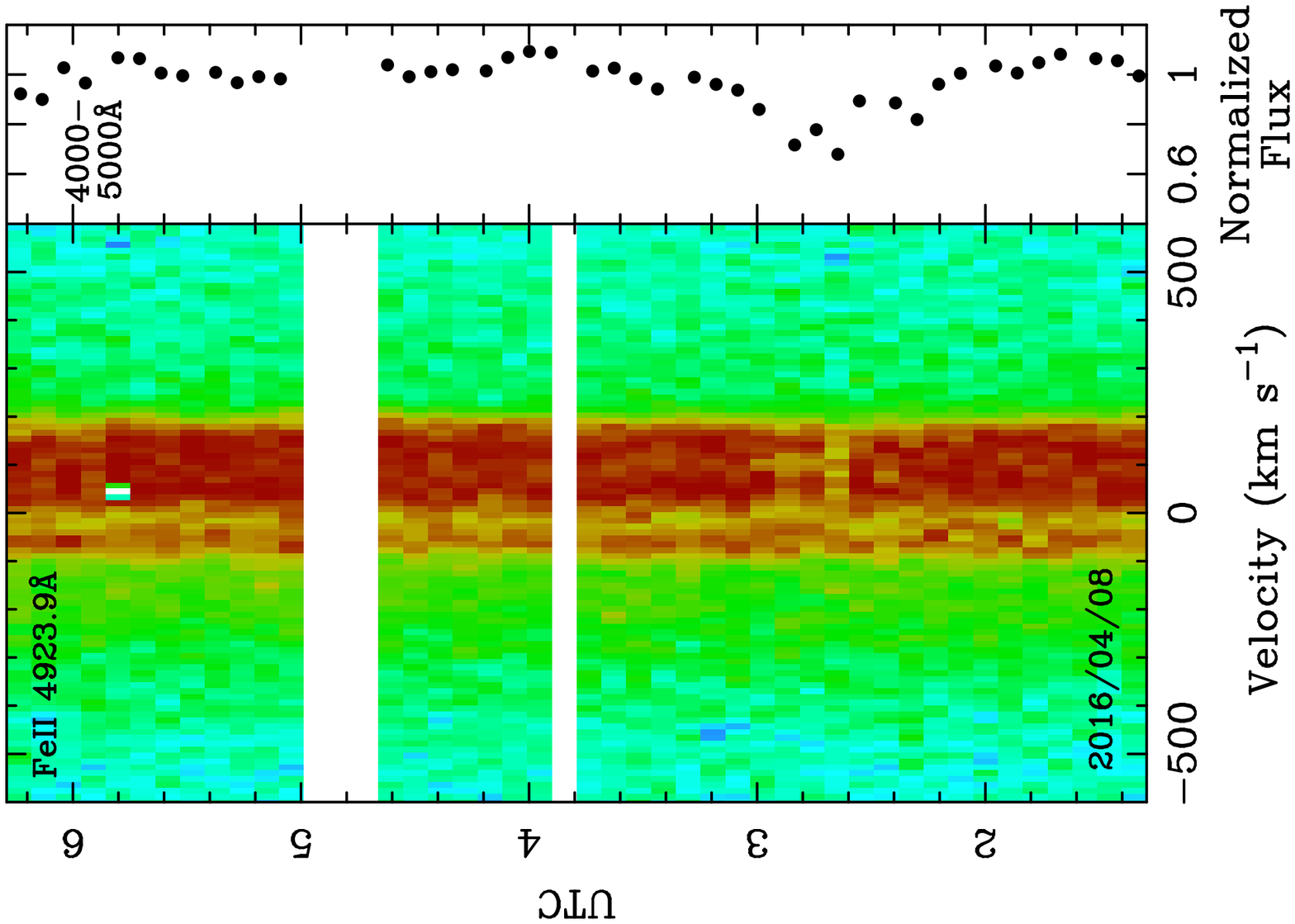}{0.47\textwidth}{}}
\caption{Running comparison of the \ion{Fe}{2} 4923.9 \AA\ spectral region.  Momentary (i.e., 10s of minutes) departures from the mean profile are detected in both the 2016 Mar and Apr observations. Broadband (4000--5000 \AA) spectrophotometry is included in the narrow right panel for each epoch.  The circumstellar line variability is clearly associated with the photometric transit events that have been seen on $\approx$ 4.5 hour periods.  \label{fig:1night}}
\end{figure*}

Our most recent observations also provide clear evidence of circumstellar variation on short time scales (i.e., tens of minutes).  Our most recent epochs include coverage of between three and five hours (slightly more than the orbital period of the transiting objects), and momentary changes in the line profile are detected most dramatically in the 2016 Mar observations.  Figure~\ref{fig:1night} show time-resolved spectra of the \ion{Fe}{2} 4923.9 \AA\ line for all three most recent epochs.  Short term variability is clearly detected in 2016 Mar over a period of approximately 20 minutes, roughly 10\% of the entire observation.  The maximum velocity remains the same, but there is a dramatic decrease in the absorption on the red side of the absorption profile.  This momentary variation coincides with a marked decrease in the spectrophotometric flux summed across all wavelengths, and corresponds in time to the most significant photometric transit signatures \citep{gaensicke16, rappaport16}.  This implies that the absorbing gas is closer to the white dwarf than the obscuring, transiting objects.  Our exposure times are significantly longer than those for the photometric monitoring, but the variation does persist over a few explosures (i.e., 10s of minutes).  It is worth noting the rarity of such events, and that for much of the observed period, the circumstellar absorption profile remained remarkably constant.

\section{A Model of the Circumstellar Environment of WD\,1145+017}

We explore three model possibilities to explain the spectral signature of the circumstellar material: (1) discrete streams, either magnetically or gravitationally formed \citep[e.g.,][]{blinova16}; (2) a curtain of magnetospheric accretion \citep{bouvier07, warner04} with a disk wind \citep{kwan07}; (3) an eccentric or warped disk \citep{metzger12, gaensicke08}.

The accretion streams seem unlikely given that we do not see much short term variability.  Even for modestly magnetic stars, these are likely narrow features that would rotate with the star.  The rotation period of the star (typically a few days; \citealt{greiss14}) is likely too long to be smeared over a single exposure.  So, due to the near-constancy of the absorption over five hours, with only short departures from the mean absorption profile, this configuration is unlikely.  In addition, we do not expect that this system is unstable to magnetic Rayleigh-Taylor instabilities \citep{blinova16}, given that the corotation radius is well beyond reasonable estimates of the magnetospheric radius.

Magnetospheric accretion models are routinely used to describe mass accretion onto pre-main sequence stars \citep[e.g.,][]{hartmann94, muzerolle04, fischer08} and have also been explored for white dwarfs \citep{warner04, patterson94}.  This model is attractive since the nearly edge-on orientation of
the WD\,1145+017 system results in an abrupt red-shifted velocity cutoff where the accreting material
flows above and below the limits of the stellar disk, a feature that is seen in all of the
circumstellar lines. Disk winds launched from the interaction region of the stellar magnetic field
and the gas disk are also a common feature of magnetospheric accretion \citep[e.g.,][]{kwan07, kurosawa16} and could be producing the narrow blue-shifted
absorption seen in some of the spectra.  We have explored model absorption line profiles produced by
an aligned dipolar magnetospheric accretion flow and find that the observed line profiles are inconsistent with this
geometry. The inconsistency is mainly due to the difficulty of reproducing the sharp red velocity
edge.  While we expect magnetospheric accretion to occur in this system and is a likely mechanism for accretion onto the star, it cannot solely account for the observed profiles.

Finally, we have constructed an eccentric disk model to explain the observed line profiles. The disk is
composed of slightly inclined confocal ellipses that extend from an inner to an outer radius. We
assume that the density is azimuthally uniform and decreases linearly with radius. The absorption
profiles are approximated as Doppler-broadened delta functions with an intrinsic line broadening of
15 \kms.  The vertical thickness of the disk is 0.1 $R_*$. The stellar parameters are taken from \citet{vanderburg15}.

\begin{figure*}[!t]
\plotone{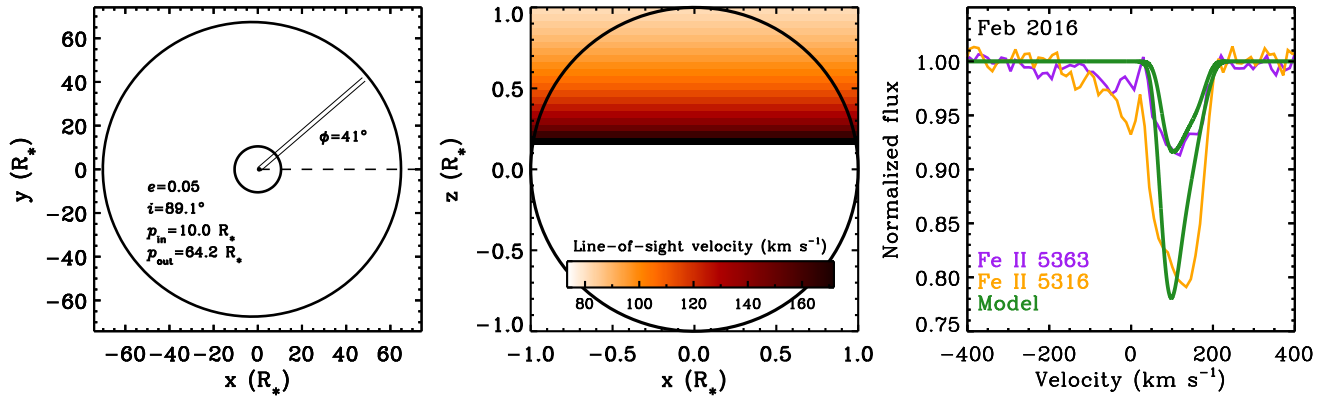}
\caption{Model of a low eccentricity disk surrounding WD\,1145+017 that successfully reproduces the observed profile shape and line ratios.  {\it (left)} Geometry of the disk.  {\it (middle)} Projected stellar disk with shading to indicate the radial velocity of disk.  Note the disk is slightly inclined.  The observed asymmetric velocity profile can be produced with a line of sight that near the apsis of the orbit.  {\it (right)}  Comparison of the model to two \ion{Fe}{2} lines showing that the profile shape and line ratios are well characterized.    
\label{fig:model}}
\end{figure*}

The model profiles for two \ion{Fe}{2} lines observed in 2016 Feb, with different oscillator strengths are
shown in Figure~\ref{fig:model} (right panel, thick green lines). We find that a slightly inclined, mildly
eccentric disk viewed at 41$^\circ$ from the apsidal line is able to roughly reproduce the
velocity range and depth of the observed red-shifted absorption. Note that there is a degeneracy between the viewing angle and the eccentricity, and we have chosen to demonstrate this model with a relatively low eccentricity disk.  We do not attempt to model the
blue-shifted absorption. We note that the size of our favored disk (see Figure~\ref{fig:model}) has similar dimensions
to that derived by \citet{melis10} for SDSS\,J104341.53+085558.2.  The truncation radius of $\approx 10 R_\star$ may be dictated by the magnetosphere \citep{bouvier07}, with reasonable values for  magnetic field (e.g., 130 G) and accretion rate (e.g., $10^{10}$ g~s$^{-1}$).  Note that current observations are only sensitive to magnetic field strengths $\gtrsim$ a few kG \citep{landstreet16}, and therefore this field would go undetected.  The total \ion{Fe}{2} mass in the model disk
is 1.1$\times$10$^{14}$ g (a minimum value given that even a modest disk inclination will result in only part of the disk is project on the stellar disk), suggesting a total disk mass less than the mass of Ceres ($M_{\rm Ceres} \approx 10^{24}$ g).  This value is in agreement with the
mass as inferred from the photospheric lines in Section~\ref{sec:photospheric}.

\subsection{Explaining the Variability}

A simple eccentric disk model is capable of explaining both the long and short term variability that is observed in the circumstellar absorption.  The monthly variability has two components.  The changes in the absorption depth can be explained by density changes of a factor of 3--8$\times$ the nominal level.  As we know that larger transiting bodies are evolving and presumably disintegrating, these density variations could be a result of the gradual feeding of material to the disk by the bodies being disrupted.  The other long term variability that is seen is the absorption profile to the blue of the photospheric line.  This feature is not accounted for in our eccentric disk model, but as discussed above, could be due to a disk wind associated with magnetospheric accretion.  Changes in disk properties and accretion rates could account for changes in the disk wind absorption strength.  The momentary variability that is seen in the 2016 Mar and 2016 Apr observations are coincident with a decrease in flux at all wavelengths, similar to the transit signatures detected in the photometric campaigns.  The spectral variability, notably the reduced absorption at the highest velocities, can be explained by a momentary shielding of the disk by an opaque, transiting body beyond the gas disk.

\section{Conclusions}

We present five epochs of high resolution and high signal-to-noise spectroscopy of the white dwarf WD\,1145+017, discovered to exhibit transiting signals of orbiting material.  The most significant conclusions are as follows:

\begin{enumerate}
\item We detect variation in the WD\,1145+017 circumstellar gas disk absorption over time scales ranging from minutes to months.  A systematic increase in equivalent width is measured over the course of the year from 2015 Apr to 2016 Apr.  In addition, significant changes in the profile shape are seen, notably a significant increase in the absorption depth on the red side of the feature and the disappearance of absorption on the blue side of the line seen in the earliest epochs.
\item Circumstellar absorption is measured in more than 250 transitions for 14 different ions of ten different elements.  We fit the circumstellar absorption with a simple trapezoidal profile and detect significant long-term variability across a period of a year.  Detected absorption arises from transitions with low excitation energies, the vast majority from $<$ 4 eV.  A correlation between equivalent width and oscillator strength is found and indicates that the absorption is not strongly optically thick (median $\tau \approx 2$).  
\item Short term variability, on the timescale of minutes, is detected in two of the three most recent epochs.  WD\,1145+017 is monitored for over 5 hours each time and momentary departures from the stable absorption feature are detected.  
\item We present a model of the circumstellar gas disk that involves an eccentric disk and magnetospheric accretion to explain the broad profile and modest absorption depth.  
\end{enumerate}

This work is an initial look at a rich data set on a unique target, as at this time, WD\,1145+017 remains the only known white dwarf with detected transit signatures.  Future work will involve a comprehensive chemical inventory of the both the accreted and circumstellar material.  The detection of \ion{O}{1} is noteworthy.  High $S/N$ observations could potentially lead to detections or stringent upper limits on hydrogen, putting interesting constraints on the volatile content.  In addition, the spectroscopic monitoring of the accretion should be correlated and compared to the detected photometric variability.  WD\,1145+017 offers the unprecedented opportunity to investigate in detail the disruption of an exoplanetesimal in real time, to constrain models of the physical and temporal processes; the production of dust and gas, and the eventual accretion onto the star.


\acknowledgments

Some of the data presented herein were obtained at the W.M. Keck Observatory from telescope time allocated to the National Aeronautics and Space Administration through the agency's scientific partnership with the California Institute of Technology and the University of California. This work was supported by a NASA Keck PI Data Award, administered by the NASA Exoplanet Science Institute. The Observatory was made possible by the generous financial support of the W.M. Keck Foundation. The authors wish to recognize and acknowledge the very significant cultural role and reverence that the summit of Mauna Kea has always had within the indigenous Hawaiian community. We are most fortunate to have the opportunity to conduct observations from this mountain.  Some of the data presented herein were made with ESO Telescopes at the La Silla Paranal Observatory under programme ID 296.C-5014.  The authors wish to thank Detlev Koester for generating the white dwarf spectral models used in this work and for his other helpful contributions to the project.  S.G.P. acknowledges financial support from the European Research Council under the European Union's Seventh Framework Programme (FP/2007-2013) under ERC-2013-ADG Grant Agreement no. 340040.  The research leading to these results has received funding from the European Research Council under the European Union's Seventh Framework Programme (FP/2007-2013) / ERC Grant Agreement n. 320964 (WDTracer).  This work was completed with support from the National Science Foundation through Astronomy and Astrophysics Research Grant AST-1313268 (PI: S.R.).

\vspace{5mm}
\facilities{Keck:I (HIRES), VLT: (X-shooter)}

\software{IRAF, IDL}

\bibliographystyle{aasjournal}

\begin{thebibliography}{}
\expandafter\ifx\csname natexlab\endcsname\relax\def\natexlab#1{#1}\fi

\bibitem[{{Alonso} {et~al.}(2016){Alonso}, {Rappaport}, {Deeg}, \&
  {Palle}}]{alonso16}
{Alonso}, R., {Rappaport}, S., {Deeg}, H.~J., \& {Palle}, E. 2016, \aap, 589,
  L6

\bibitem[{{Bergfors} {et~al.}(2014){Bergfors}, {Farihi}, {Dufour}, \&
  {Rocchetto}}]{bergfors14}
{Bergfors}, C., {Farihi}, J., {Dufour}, P., \& {Rocchetto}, M. 2014, \mnras,
  444, 2147

\bibitem[{{Blinova} {et~al.}(2016){Blinova}, {Romanova}, \&
  {Lovelace}}]{blinova16}
{Blinova}, A.~A., {Romanova}, M.~M., \& {Lovelace}, R.~V.~E. 2016, \mnras, 459,
  2354

\bibitem[{{Bouvier} {et~al.}(2007){Bouvier}, {Alencar}, {Harries},
  {Johns-Krull}, \& {Romanova}}]{bouvier07}
{Bouvier}, J., {Alencar}, S.~H.~P., {Harries}, T.~J., {Johns-Krull}, C.~M., \&
  {Romanova}, M.~M. 2007, Protostars and Planets V, 479

\bibitem[{{Carson} {et~al.}(2013){Carson}, {Thalmann}, {Janson}, {Kozakis},
  {Bonnefoy}, {Biller}, {Schlieder}, {Currie}, {McElwain}, {Goto}, {Henning},
  {Brandner}, {Feldt}, {Kandori}, {Kuzuhara}, {Stevens}, {Wong}, {Gainey},
  {Fukagawa}, {Kuwada}, {Brandt}, {Kwon}, {Abe}, {Egner}, {Grady}, {Guyon},
  {Hashimoto}, {Hayano}, {Hayashi}, {Hayashi}, {Hodapp}, {Ishii}, {Iye},
  {Knapp}, {Kudo}, {Kusakabe}, {Matsuo}, {Miyama}, {Morino}, {Moro-Martin},
  {Nishimura}, {Pyo}, {Serabyn}, {Suto}, {Suzuki}, {Takami}, {Takato},
  {Terada}, {Tomono}, {Turner}, {Watanabe}, {Wisniewski}, {Yamada}, {Takami},
  {Usuda}, \& {Tamura}}]{carson13}
{Carson}, J., {Thalmann}, C., {Janson}, M., {et~al.} 2013, \apjl, 763, L32

\bibitem[{{Farihi} {et~al.}(2013{\natexlab{a}}){Farihi}, {G{\"a}nsicke}, \&
  {Koester}}]{farihi13b}
{Farihi}, J., {G{\"a}nsicke}, B.~T., \& {Koester}, D. 2013{\natexlab{a}},
  Science, 342, 218

\bibitem[{{Farihi} {et~al.}(2013{\natexlab{b}}){Farihi}, {G{\"a}nsicke}, \&
  {Koester}}]{farihi13}
---. 2013{\natexlab{b}}, \mnras, 432, 1955

\bibitem[{{Farihi} {et~al.}(2012){Farihi}, {G{\"a}nsicke}, {Wyatt}, {Girven},
  {Pringle}, \& {King}}]{farihi12}
{Farihi}, J., {G{\"a}nsicke}, B.~T., {Wyatt}, M.~C., {et~al.} 2012, \mnras,
  424, 464

\bibitem[{{Farihi} {et~al.}(2009){Farihi}, {Jura}, \& {Zuckerman}}]{farihi09}
{Farihi}, J., {Jura}, M., \& {Zuckerman}, B. 2009, \apj, 694, 805

\bibitem[{{Fischer} {et~al.}(2008){Fischer}, {Kwan}, {Edwards}, \&
  {Hillenbrand}}]{fischer08}
{Fischer}, W., {Kwan}, J., {Edwards}, S., \& {Hillenbrand}, L. 2008, \apj, 687,
  1117
  
  \bibitem[{{Frisch} {et~al.}(2011){Frisch}, {Redfield}, \& {Slavin}}]{frisch11}
{Frisch}, P.~C., {Redfield}, S., \& {Slavin}, J.~D. 2011, \araa, 49, 237

\bibitem[{{G{\"a}nsicke} {et~al.}(2012){G{\"a}nsicke}, {Koester}, {Farihi},
  {Girven}, {Parsons}, \& {Breedt}}]{gaensicke12}
{G{\"a}nsicke}, B.~T., {Koester}, D., {Farihi}, J., {et~al.} 2012, \mnras, 424,
  333

\bibitem[{{G{\"a}nsicke} {et~al.}(2008){G{\"a}nsicke}, {Koester}, {Marsh},
  {Rebassa-Mansergas}, \& {Southworth}}]{gaensicke08}
{G{\"a}nsicke}, B.~T., {Koester}, D., {Marsh}, T.~R., {Rebassa-Mansergas}, A.,
  \& {Southworth}, J. 2008, \mnras, 391, L103

\bibitem[{{G{\"a}nsicke} {et~al.}(2006){G{\"a}nsicke}, {Marsh}, {Southworth},
  \& {Rebassa-Mansergas}}]{gaensicke06}
{G{\"a}nsicke}, B.~T., {Marsh}, T.~R., {Southworth}, J., \&
  {Rebassa-Mansergas}, A. 2006, Science, 314, 1908

\bibitem[{{G{\"a}nsicke} {et~al.}(2016){G{\"a}nsicke}, {Aungwerojwit}, {Marsh},
  {Dhillon}, {Sahman}, {Veras}, {Farihi}, {Chote}, {Ashley}, {Arjyotha},
  {Rattanasoon}, {Littlefair}, {Pollacco}, \& {Burleigh}}]{gaensicke16}
{G{\"a}nsicke}, B.~T., {Aungwerojwit}, A., {Marsh}, T.~R., {et~al.} 2016,
  \apjl, 818, L7

\bibitem[{{Greiss} {et~al.}(2014){Greiss}, {G{\"a}nsicke}, {Hermes}, {Steeghs},
  {Koester}, {Ramsay}, {Barclay}, \& {Townsley}}]{greiss14}
{Greiss}, S., {G{\"a}nsicke}, B.~T., {Hermes}, J.~J., {et~al.} 2014, \mnras,
  438, 3086

\bibitem[{{Hartmann} {et~al.}(1994){Hartmann}, {Hewett}, \&
  {Calvet}}]{hartmann94}
{Hartmann}, L., {Hewett}, R., \& {Calvet}, N. 1994, \apj, 426, 669

\bibitem[{{Johnson} {et~al.}(2015){Johnson}, {Redfield}, \&
  {Jensen}}]{johnson15}
{Johnson}, M.~C., {Redfield}, S., \& {Jensen}, A.~G. 2015, \apj, 807, 162

\bibitem[{{Jura} \& {Young}(2014)}]{jura14}
{Jura}, M., \& {Young}, E.~D. 2014, Annual Review of Earth and Planetary
  Sciences, 42, 45

\bibitem[{{Kipping} {et~al.}(2014){Kipping}, {Torres}, {Buchhave}, {Kenyon},
  {Henze}, {Isaacson}, {Kolbl}, {Marcy}, {Bryson}, {Stassun}, \&
  {Bastien}}]{kipping14}
{Kipping}, D.~M., {Torres}, G., {Buchhave}, L.~A., {et~al.} 2014, \apj, 795, 25

\bibitem[{{Koester}(2009)}]{koester09}
{Koester}, D. 2009, \aap, 498, 517

\bibitem[{{Koester}(2010)}]{koester10}
---. 2010, \memsai, 81, 921

\bibitem[{{Koester} {et~al.}(2014){Koester}, {G{\"a}nsicke}, \&
  {Farihi}}]{koester14}
{Koester}, D., {G{\"a}nsicke}, B.~T., \& {Farihi}, J. 2014, \aap, 566, A34

\bibitem[{{Kurosawa} {et~al.}(2016){Kurosawa}, {Kreplin}, {Weigelt}, {Natta},
  {Benisty}, {Isella}, {Tatulli}, {Massi}, {Testi}, {Kraus}, {Duvert},
  {Petrov}, \& {Stee}}]{kurosawa16}
{Kurosawa}, R., {Kreplin}, A., {Weigelt}, G., {et~al.} 2016, \mnras, 457, 2236

\bibitem[{{Kwan} {et~al.}(2007){Kwan}, {Edwards}, \& {Fischer}}]{kwan07}
{Kwan}, J., {Edwards}, S., \& {Fischer}, W. 2007, \apj, 657, 897

\bibitem[{{Landstreet} {et~al.}(2016){Landstreet}, {Bagnulo}, {Martin}, \&
  {Valyavin}}]{landstreet16}
{Landstreet}, J.~D., {Bagnulo}, S., {Martin}, A., \& {Valyavin}, G. 2016, \aap,
  591, A80

\bibitem[{{Macintosh} {et~al.}(2015){Macintosh}, {Graham}, {Barman}, {De Rosa},
  {Konopacky}, {Marley}, {Marois}, {Nielsen}, {Pueyo}, {Rajan}, {Rameau},
  {Saumon}, {Wang}, {Patience}, {Ammons}, {Arriaga}, {Artigau}, {Beckwith},
  {Brewster}, {Bruzzone}, {Bulger}, {Burningham}, {Burrows}, {Chen}, {Chiang},
  {Chilcote}, {Dawson}, {Dong}, {Doyon}, {Draper}, {Duch{\^e}ne}, {Esposito},
  {Fabrycky}, {Fitzgerald}, {Follette}, {Fortney}, {Gerard}, {Goodsell},
  {Greenbaum}, {Hibon}, {Hinkley}, {Cotten}, {Hung}, {Ingraham},
  {Johnson-Groh}, {Kalas}, {Lafreniere}, {Larkin}, {Lee}, {Line}, {Long},
  {Maire}, {Marchis}, {Matthews}, {Max}, {Metchev}, {Millar-Blanchaer},
  {Mittal}, {Morley}, {Morzinski}, {Murray-Clay}, {Oppenheimer}, {Palmer},
  {Patel}, {Perrin}, {Poyneer}, {Rafikov}, {Rantakyr{\"o}}, {Rice}, {Rojo},
  {Rudy}, {Ruffio}, {Ruiz}, {Sadakuni}, {Saddlemyer}, {Salama}, {Savransky},
  {Schneider}, {Sivaramakrishnan}, {Song}, {Soummer}, {Thomas}, {Vasisht},
  {Wallace}, {Ward-Duong}, {Wiktorowicz}, {Wolff}, \&
  {Zuckerman}}]{macintosh15}
{Macintosh}, B., {Graham}, J.~R., {Barman}, T., {et~al.} 2015, Science, 350, 64

\bibitem[{{Madhusudhan} {et~al.}(2012){Madhusudhan}, {Lee}, \&
  {Mousis}}]{madhu12b}
{Madhusudhan}, N., {Lee}, K.~K.~M., \& {Mousis}, O. 2012, \apjl, 759, L40

\bibitem[{{Madhusudhan} \& {Redfield}(2015)}]{madhu15}
{Madhusudhan}, N., \& {Redfield}, S. 2015, International Journal of
  Astrobiology, 14, 177

\bibitem[{{Manser} {et~al.}(2016){Manser}, {G{\"a}nsicke}, {Marsh}, {Veras},
  {Koester}, {Breedt}, {Pala}, {Parsons}, \& {Southworth}}]{manser16}
{Manser}, C.~J., {G{\"a}nsicke}, B.~T., {Marsh}, T.~R., {et~al.} 2016, \mnras,
  455, 4467

\bibitem[{{Marois} {et~al.}(2008){Marois}, {Macintosh}, {Barman}, {Zuckerman},
  {Song}, {Patience}, {Lafreni{\`e}re}, \& {Doyon}}]{marois08}
{Marois}, C., {Macintosh}, B., {Barman}, T., {et~al.} 2008, Science, 322, 1348

\bibitem[{{Melis} {et~al.}(2010){Melis}, {Jura}, {Albert}, {Klein}, \&
  {Zuckerman}}]{melis10}
{Melis}, C., {Jura}, M., {Albert}, L., {Klein}, B., \& {Zuckerman}, B. 2010,
  \apj, 722, 1078

\bibitem[{{Metzger} {et~al.}(2012){Metzger}, {Rafikov}, \&
  {Bochkarev}}]{metzger12}
{Metzger}, B.~D., {Rafikov}, R.~R., \& {Bochkarev}, K.~V. 2012, \mnras, 423,
  505

\bibitem[{{Miller-Ricci} \& {Fortney}(2010)}]{millerricci10}
{Miller-Ricci}, E., \& {Fortney}, J.~J. 2010, \apjl, 716, L74

\bibitem[{{Muzerolle} {et~al.}(2004){Muzerolle}, {D'Alessio}, {Calvet}, \&
  {Hartmann}}]{muzerolle04}
{Muzerolle}, J., {D'Alessio}, P., {Calvet}, N., \& {Hartmann}, L. 2004, \apj,
  617, 406

\bibitem[{{{\"O}berg} {et~al.}(2015){{\"O}berg}, {Guzm{\'a}n}, {Furuya}, {Qi},
  {Aikawa}, {Andrews}, {Loomis}, \& {Wilner}}]{oberg15}
{{\"O}berg}, K.~I., {Guzm{\'a}n}, V.~V., {Furuya}, K., {et~al.} 2015, \nat,
  520, 198

\bibitem[{{Patterson}(1994)}]{patterson94}
{Patterson}, J. 1994, \pasp, 106, 209

\bibitem[{{Raddi} {et~al.}(2015){Raddi}, {G{\"a}nsicke}, {Koester}, {Farihi},
  {Hermes}, {Scaringi}, {Breedt}, \& {Girven}}]{raddi15}
{Raddi}, R., {G{\"a}nsicke}, B.~T., {Koester}, D., {et~al.} 2015, \mnras, 450,
  2083

\bibitem[{{Rappaport} {et~al.}(2016){Rappaport}, {Gary}, {Kaye}, {Vanderburg},
  {Croll}, {Benni}, \& {Foote}}]{rappaport16}
{Rappaport}, S., {Gary}, B.~L., {Kaye}, T., {et~al.} 2016, \mnras,
  arXiv:1602.00740

\bibitem[{{Redfield} {et~al.}(2008){Redfield}, {Endl}, {Cochran}, \&
  {Koesterke}}]{redfield08brtr}
{Redfield}, S., {Endl}, M., {Cochran}, W.~D., \& {Koesterke}, L. 2008, \apjl,
  673, L87

\bibitem[{{Rocchetto} {et~al.}(2015){Rocchetto}, {Farihi}, {G{\"a}nsicke}, \&
  {Bergfors}}]{rocchetto15}
{Rocchetto}, M., {Farihi}, J., {G{\"a}nsicke}, B.~T., \& {Bergfors}, C. 2015,
  \mnras, 449, 574

\bibitem[{{Su} {et~al.}(2013){Su}, {Rieke}, {Malhotra}, {Stapelfeldt},
  {Hughes}, {Bonsor}, {Wilner}, {Balog}, {Watson}, {Werner}, \&
  {Misselt}}]{su13}
{Su}, K.~Y.~L., {Rieke}, G.~H., {Malhotra}, R., {et~al.} 2013, \apj, 763, 118

\bibitem[{{van der Marel} {et~al.}(2013){van der Marel}, {van Dishoeck},
  {Bruderer}, {Birnstiel}, {Pinilla}, {Dullemond}, {van Kempen}, {Schmalzl},
  {Brown}, {Herczeg}, {Mathews}, \& {Geers}}]{vandermarel13}
{van der Marel}, N., {van Dishoeck}, E.~F., {Bruderer}, S., {et~al.} 2013,
  Science, 340, 1199

\bibitem[{{Vanderburg} {et~al.}(2015){Vanderburg}, {Johnson}, {Rappaport},
  {Bieryla}, {Irwin}, {Lewis}, {Kipping}, {Brown}, {Dufour}, {Ciardi}, {Angus},
  {Schaefer}, {Latham}, {Charbonneau}, {Beichman}, {Eastman}, {McCrady},
  {Wittenmyer}, \& {Wright}}]{vanderburg15}
{Vanderburg}, A., {Johnson}, J.~A., {Rappaport}, S., {et~al.} 2015, \nat, 526,
  546

\bibitem[{{Vanderburg} {et~al.}(2016){Vanderburg}, {Latham}, {Buchhave},
  {Bieryla}, {Berlind}, {Calkins}, {Esquerdo}, {Welsh}, \&
  {Johnson}}]{vanderburg16}
{Vanderburg}, A., {Latham}, D.~W., {Buchhave}, L.~A., {et~al.} 2016, \apjs,
  222, 14

\bibitem[{{Veras} {et~al.}(2015){Veras}, {Leinhardt}, {Eggl}, \&
  {G{\"a}nsicke}}]{veras15}
{Veras}, D., {Leinhardt}, Z.~M., {Eggl}, S., \& {G{\"a}nsicke}, B.~T. 2015,
  \mnras, 451, 3453

\bibitem[{{Vernet} {et~al.}(2011){Vernet}, {Dekker}, {D'Odorico}, {Kaper},
  {Kjaergaard}, {Hammer}, {Randich}, {Zerbi}, {Groot}, {Hjorth}, {Guinouard},
  {Navarro}, {Adolfse}, {Albers}, {Amans}, {Andersen}, {Andersen}, {Binetruy},
  {Bristow}, {Castillo}, {Chemla}, {Christensen}, {Conconi}, {Conzelmann},
  {Dam}, {de Caprio}, {de Ugarte Postigo}, {Delabre}, {di Marcantonio},
  {Downing}, {Elswijk}, {Finger}, {Fischer}, {Flores}, {Fran{\c c}ois},
  {Goldoni}, {Guglielmi}, {Haigron}, {Hanenburg}, {Hendriks}, {Horrobin},
  {Horville}, {Jessen}, {Kerber}, {Kern}, {Kiekebusch}, {Kleszcz}, {Klougart},
  {Kragt}, {Larsen}, {Lizon}, {Lucuix}, {Mainieri}, {Manuputy}, {Martayan},
  {Mason}, {Mazzoleni}, {Michaelsen}, {Modigliani}, {Moehler}, {M{\o}ller},
  {Norup S{\o}rensen}, {N{\o}rregaard}, {P{\'e}roux}, {Patat}, {Pena}, {Pragt},
  {Reinero}, {Rigal}, {Riva}, {Roelfsema}, {Royer}, {Sacco}, {Santin},
  {Schoenmaker}, {Spano}, {Sweers}, {Ter Horst}, {Tintori}, {Tromp}, {van
  Dael}, {van der Vliet}, {Venema}, {Vidali}, {Vinther}, {Vola}, {Winters},
  {Wistisen}, {Wulterkens}, \& {Zacchei}}]{vernet11}
{Vernet}, J., {Dekker}, H., {D'Odorico}, S., {et~al.} 2011, \aap, 536, A105

\bibitem[{{Vogt} {et~al.}(1994){Vogt}, {Allen}, {Bigelow}, {Bresee}, {Brown},
  {Cantrall}, {Conrad}, {Couture}, {Delaney}, {Epps}, {Hilyard}, {Hilyard},
  {Horn}, {Jern}, {Kanto}, {Keane}, {Kibrick}, {Lewis}, {Osborne},
  {Pardeilhan}, {Pfister}, {Ricketts}, {Robinson}, {Stover}, {Tucker}, {Ward},
  \& {Wei}}]{vogt94}
{Vogt}, S.~S., {Allen}, S.~L., {Bigelow}, B.~C., {et~al.} 1994, in \procspie,
  Vol. 2198, Instrumentation in Astronomy VIII, ed. D.~L. {Crawford} \& E.~R.
  {Craine}, 362

\bibitem[{{Warner}(2004)}]{warner04}
{Warner}, B. 2004, \pasp, 116, 115

\bibitem[{{Wilson} {et~al.}(2014){Wilson}, {G{\"a}nsicke}, {Koester}, {Raddi},
  {Breedt}, {Southworth}, \& {Parsons}}]{wilson14}
{Wilson}, D.~J., {G{\"a}nsicke}, B.~T., {Koester}, D., {et~al.} 2014, \mnras,
  445, 1878

\bibitem[{{Wyatt} {et~al.}(2014){Wyatt}, {Farihi}, {Pringle}, \&
  {Bonsor}}]{wyatt14}
{Wyatt}, M.~C., {Farihi}, J., {Pringle}, J.~E., \& {Bonsor}, A. 2014, \mnras,
  439, 3371

\bibitem[{{Xu} \& {Jura}(2012)}]{xu12}
{Xu}, S., \& {Jura}, M. 2012, \apj, 745, 88

\bibitem[{{Xu} \& {Jura}(2014)}]{xu14}
---. 2014, \apjl, 792, L39

\bibitem[{{Xu} {et~al.}(2016){Xu}, {Jura}, {Dufour}, \& {Zuckerman}}]{xu16}
{Xu}, S., {Jura}, M., {Dufour}, P., \& {Zuckerman}, B. 2016, \apjl, 816, L22

\bibitem[{{Xu} {et~al.}(2013){Xu}, {Jura}, {Klein}, {Koester}, \&
  {Zuckerman}}]{xu13}
{Xu}, S., {Jura}, M., {Klein}, B., {Koester}, D., \& {Zuckerman}, B. 2013,
  \apj, 766, 132

\bibitem[{{Zuckerman} {et~al.}(2010){Zuckerman}, {Melis}, {Klein}, {Koester},
  \& {Jura}}]{zuckerman10}
{Zuckerman}, B., {Melis}, C., {Klein}, B., {Koester}, D., \& {Jura}, M. 2010,
  \apj, 722, 725

\end{thebibliography}

\pagebreak
\floattable
\begin{deluxetable*}{rlRR@{\extracolsep{3pt}}RCCCCCc}
\tablecaption{Detected Circumstellar Absorption Features and Fit Parameters \label{tab:table}}
\tablehead{
\colhead{Ion} & \colhead{$\lambda$} & \colhead{$E_{\rm low}$} & \colhead{$E_{\rm up}$} & \colhead{$\log gf$} & \colhead{EW} & \colhead{EW} & \colhead{EW} & \colhead{EW} & \colhead{EW} & \colhead{Comments\tablenotemark{a}} \\
\colhead{} & \colhead{} & \colhead{} & \colhead{} & \colhead{} & \multicolumn{2}{c}{Keck / HIRES} & \multicolumn{3}{c}{VLT / X-shooter} & \colhead{} \\ \cline{6-7} \cline{8-10}
\colhead{} & \colhead{} & \colhead{} & \colhead{} & \colhead{} & \colhead{2015 Apr} & \colhead{2015 Nov} & \colhead{2016 Feb} & \colhead{2016 Mar} & \colhead{2016 Apr} & \colhead{} \\
\colhead{} & \colhead{(\AA)} & \colhead{(eV)} & \colhead{(eV)} & \colhead{} & \colhead{(\AA)} & \colhead{(\AA)} & \colhead{(\AA)} & \colhead{(\AA)} & \colhead{(\AA)} & \colhead{}
}
\startdata
 Fe II & 3227.742 &  1.67 &  5.51 & -1.130 & 0.328 \pm ^{0.031}_{0.030} & 0.325 \pm ^{0.035}_{0.035} & 0.596 \pm ^{0.047}_{0.043} & 0.698 \pm ^{0.026}_{0.032} & 0.768 \pm ^{0.017}_{0.035} & i \\
 Ni II & 3576.764 &  3.07 &  6.54 & -1.628 & 0.181 \pm ^{0.015}_{0.014} & 0.197 \pm ^{0.026}_{0.021} & 0.501 \pm ^{0.021}_{0.018} & 0.545 \pm ^{0.021}_{0.017} & 0.549 \pm ^{0.010}_{0.013} & i \\
  Fe I & 3647.842 &  0.91 &  4.31 & -0.194 & 0.026 \pm ^{0.004}_{0.004} & 0.571 \pm ^{0.033}_{0.034} & 0.083 \pm ^{0.017}_{0.016} & 0.098 \pm ^{0.012}_{0.011} & 0.140 \pm ^{0.015}_{0.007} & i \\
 Ca II & 3933.663 &  0.00 &  3.15 &  0.134 & 0.417 \pm ^{0.015}_{0.015} & 0.395 \pm ^{0.023}_{0.023} & 0.477 \pm ^{0.015}_{0.013} & 0.462 \pm ^{0.010}_{0.008} & 0.542 \pm ^{0.006}_{0.006} & i \\
 Ca II & 3968.469 &  0.00 &  3.12 & -0.166 & 0.395 \pm ^{0.024}_{0.023} & 0.366 \pm ^{0.018}_{0.017} & 0.393 \pm ^{0.046}_{0.011} & 0.520 \pm ^{0.006}_{0.006} & 0.614 \pm ^{0.005}_{0.005} & i \\
  Fe I & 4045.813 &  1.48 &  4.55 &  0.280 & \nodata & \nodata & 0.107 \pm ^{0.010}_{0.009} & 0.151 \pm ^{0.011}_{0.013} & 0.187 \pm ^{0.006}_{0.006} & i \\
 Fe II & 4173.461 &  2.58 &  5.55 & -2.180 & 0.101 \pm ^{0.013}_{0.012} & 0.115 \pm ^{0.013}_{0.013} & 0.165 \pm ^{0.008}_{0.007} & 0.169 \pm ^{0.009}_{0.007} & 0.199 \pm ^{0.021}_{0.005} & i \\
 Fe II & 4178.862 &  2.58 &  5.55 & -2.480 & 0.111 \pm ^{0.012}_{0.011} & 0.161 \pm ^{0.012}_{0.012} & 0.236 \pm ^{0.011}_{0.009} & 0.269 \pm ^{0.025}_{0.009} & 0.265 \pm ^{0.005}_{0.008} & i \\
 Fe II & 4233.172 &  2.58 &  5.51 & -2.000 & 0.272 \pm ^{0.009}_{0.007} & 0.330 \pm ^{0.014}_{0.013} & 0.584 \pm ^{0.026}_{0.041} & 0.662 \pm ^{0.016}_{0.007} & 0.721 \pm ^{0.004}_{0.005} & i \\
  Fe I & 4325.762 &  1.61 &  4.47 & -0.010 & 0.021 \pm ^{0.005}_{0.005} & 0.057 \pm ^{0.014}_{0.013} & 0.117 \pm ^{0.010}_{0.012} & 0.152 \pm ^{0.009}_{0.008} & 0.183 \pm ^{0.006}_{0.005} & i \\
 Fe II & 4351.769 &  2.70 &  5.55 & -2.100 & 0.213 \pm ^{0.008}_{0.008} & 0.221 \pm ^{0.014}_{0.013} & 0.416 \pm ^{0.009}_{0.010} & 0.511 \pm ^{0.006}_{0.007} & 0.546 \pm ^{0.004}_{0.004} & i \\
  Fe I & 4404.750 &  1.56 &  4.37 & -0.142 & 0.029 \pm ^{0.009}_{0.008} & 0.022 \pm ^{0.008}_{0.007} & 0.056 \pm ^{0.008}_{0.007} & 0.084 \pm ^{0.007}_{0.006} & 0.104 \pm ^{0.004}_{0.005} & i \\
   Fe II & 4508.288 &  2.86 &  5.60 & -2.210 & 0.128 \pm ^{0.012}_{0.011} & 0.203 \pm ^{0.018}_{0.017} & 0.267 \pm ^{0.018}_{0.012} & 0.325 \pm ^{0.010}_{0.019} & 0.338 \pm ^{0.005}_{0.004} & i \\
Ti II & 4533.969 &  1.24 &  3.97 & -0.770 & 0.454 \pm ^{0.784}_{0.441} & 0.612 \pm ^{1.109}_{0.586} & 0.041 \pm ^{0.010}_{0.008} & 0.087 \pm ^{0.009}_{0.009} & 0.097 \pm ^{0.006}_{0.006} & i \\
 Fe II & 4549.474 &  2.83 &  5.55 & -1.750 & 0.251 \pm ^{0.017}_{0.016} & 0.304 \pm ^{0.016}_{0.016} & 0.403 \pm ^{0.023}_{0.017} & 0.512 \pm ^{0.012}_{0.012} & 0.574 \pm ^{0.009}_{0.016} & i \\
 Fe II & 4583.837 &  2.81 &  5.51 & -2.020 & 0.289 \pm ^{0.008}_{0.008} & 0.409 \pm ^{0.018}_{0.017} & 0.610 \pm ^{0.021}_{0.013} & 0.688 \pm ^{0.007}_{0.007} & 0.745 \pm ^{0.004}_{0.004} & i \\
 Fe II & 4629.339 &  2.81 &  5.48 & -2.370 & 0.119 \pm ^{0.008}_{0.008} & 0.174 \pm ^{0.015}_{0.015} & 0.267 \pm ^{0.020}_{0.014} & 0.290 \pm ^{0.028}_{0.010} & 0.324 \pm ^{0.004}_{0.004} & i \\
Cr II & 4824.127 &  3.87 &  6.44 & -1.220 & 0.025 \pm ^{1.047}_{0.013} & 0.030 \pm ^{0.010}_{0.011} & 0.049 \pm ^{0.015}_{0.013} & 0.070 \pm ^{0.011}_{0.009} & 0.091 \pm ^{0.013}_{0.010} & i \\
 Fe II & 4923.927 &  2.89 &  5.41 & -1.320 & 0.459 \pm ^{0.027}_{0.024} & 0.688 \pm ^{0.033}_{0.031} & 1.182 \pm ^{0.027}_{0.025} & 1.271 \pm ^{0.015}_{0.012} & 1.402 \pm ^{0.008}_{0.031} & i \\
  Mg I & 5183.604 &  2.72 &  5.11 & -0.180 & 0.070 \pm ^{0.008}_{0.008} & 0.159 \pm ^{0.035}_{0.035} & 0.147 \pm ^{0.012}_{0.012} & 0.179 \pm ^{0.012}_{0.013} & 0.188 \pm ^{0.007}_{0.007} & i \\
 Fe II & 5234.625 &  3.22 &  5.59 & -2.050 & 0.173 \pm ^{0.011}_{0.010} & 0.241 \pm ^{0.026}_{0.025} & 0.397 \pm ^{0.011}_{0.013} & 0.492 \pm ^{0.009}_{0.009} & 0.512 \pm ^{0.004}_{0.005} & i \\
 Fe II & 5276.002 &  3.20 &  5.55 & -1.940 & 0.239 \pm ^{0.012}_{0.013} & 0.342 \pm ^{0.044}_{0.044} & 0.472 \pm ^{0.012}_{0.011} & 0.533 \pm ^{0.033}_{0.011} & 0.583 \pm ^{0.012}_{0.016} & i \\
 Fe II & 5316.615 &  3.15 &  5.48 & -1.850 & 0.313 \pm ^{0.012}_{0.011} & 0.522 \pm ^{0.033}_{0.030} & 0.654 \pm ^{0.022}_{0.018} & 0.750 \pm ^{0.025}_{0.013} & 0.882 \pm ^{0.005}_{0.005} & i \\
 Fe II & 5362.869 &  3.20 &  5.51 & -2.739 & 0.143 \pm ^{0.011}_{0.010} & 0.100 \pm ^{0.021}_{0.019} & 0.246 \pm ^{0.016}_{0.014} & 0.304 \pm ^{0.012}_{0.011} & 0.308 \pm ^{0.010}_{0.007} & i \\
  Na I & 5889.951 &  0.00 &  2.10 &  0.117 & \nodata & \nodata & 0.297 \pm ^{0.028}_{0.039} & 0.347 \pm ^{0.022}_{0.020} & 0.408 \pm ^{0.016}_{0.026} & i \\
  Na I & 5895.924 &  0.00 &  2.10 & -0.184 & \nodata & \nodata & 0.625 \pm ^{0.051}_{0.055} & 0.692 \pm ^{0.043}_{0.039} & 0.977 \pm ^{0.024}_{0.020} & b \\
 Fe II & 6247.557 &  3.89 &  5.88 & -2.329 & \nodata & \nodata & 0.158 \pm ^{0.014}_{0.014} & 0.244 \pm ^{0.017}_{0.015} & 0.302 \pm ^{0.011}_{0.010} & i \\
 Fe II & 6456.383 &  3.90 &  5.82 & -2.075 & \nodata & \nodata & 0.350 \pm ^{0.019}_{0.017} & 0.374 \pm ^{0.024}_{0.023} & 0.491 \pm ^{0.018}_{0.027} & i \\
   O I & 7775.388 &  9.15 & 10.74 & -0.046 & \nodata & \nodata & 0.351 \pm ^{0.024}_{0.021} & 0.443 \pm ^{0.024}_{0.022} & 0.475 \pm ^{0.014}_{0.014} & i \\
 Ca II & 8498.023 &  1.69 &  3.15 & -1.312 & \nodata & \nodata & 0.366 \pm ^{0.036}_{0.040} & 0.453 \pm ^{0.024}_{0.025} & 0.553 \pm ^{0.014}_{0.014} & i \\
 Ca II & 8542.091 &  1.70 &  3.15 & -0.362 & \nodata & \nodata & 0.650 \pm ^{0.024}_{0.029} & 0.732 \pm ^{0.029}_{0.024} & 0.801 \pm ^{0.017}_{0.044} & i \\
 Ca II & 8662.141 &  1.69 &  3.12 & -0.623 & \nodata & \nodata & 0.533 \pm ^{0.025}_{0.024} & 0.763 \pm ^{0.035}_{0.044} & 0.766 \pm ^{0.017}_{0.016} & i \\
 \enddata
 \tablenotetext{a}{(i) isolated feature; (b) blended feature}
\tablecomments{Table \ref{tab:table} is published 
in its entirety in the machine readable format.  A portion is
shown here for guidance regarding its form and content.}
\end{deluxetable*}


\end{document}